# Calculation of Binding Energies for Fractional Quantum Hall States with Even Denominators

Shosuke Sasaki

*Shizuoka Institute of Science and Technology 2200-2 Toyosawa Fukuroi, 437-8555, Japan*

Fractional quantum Hall states with even denominators have the following specific properties: states with filling factors $\nu=5/8$, $7/10$, $3/8$, $3/10$, and so on have respective local minima in the experimental curve of diagonal resistivity $\rho_{xx}$ versus magnetic field strength. These states are not standard composite fermion states and are described in the expanded framework. For that reason, the binding energies of these states are not obtained. Therefore, it is meaningful to calculate those binding energies using various means. We calculate the binding energies of electron pairs in nearest neighbor orbitals or nearest neighbor hole pairs using the second-order perturbation method for the Coulomb interactions among many electrons. The calculated binding energies per electron are $(1/10)Z_2$ for $\nu=5/8$, $(2/35)Z_2$ for $\nu=7/10$, $(1/6)Z_2$ for $\nu=3/8$ and $(2/15)Z_2$ for $\nu=3/10$ and so on, but they are zero for $\nu=1/2$, $\nu=1/4$, $\nu=3/4$, $\nu=1/6$, $\nu=5/6$, $\nu=1/8$, $\nu=7/8$, $\nu=1/10$, $\nu=9/10$, $\nu=1/12$ and $\nu=11/12$. The higher order calculations also show the same behavior as in the second order. These results further elucidate some aspects of experimental data.

## 1. Introduction

Precise experiments for the fractional quantum Hall effect (FQHE) have been carried out in ultra-high-mobility samples.[1,2] Many local minima of diagonal resistivity $\rho_{xx}$ have been discovered. The local minima are detected at the filling factors $\nu=3/8$, $3/10$, $4/11$, $4/13$, $5/13$, $5/17$, $6/17$, and so on. These states cannot be understood using the standard composite fermion (CF) model.[3] Wojs et al.[4], Smet[5] and Pashitskii[6] investigated these states and described the states in their extended systematics. Particularly, Smet has explained these states in terms of multiflavor composite fermion pictures with coexistence of composite fermions carrying different numbers of fluxes. Pashitskii presented expanded systematics based on Halperin's conjecture of coexistence of free electrons and bound electron pairs, with predicted new exotic fractions of ν=5/14, 5/16, 3/20. Subsequently, several investigations for the FQH states have been undertaken with even denominators. However, their binding energies are not obtained. Moreover, many fractional filling factors are classifiable into three categories as original Laughlin states (ν=1/3, 1/5, 1/7, ••••), CF states (ν=2/5, 3/7, •••, 2/9, 3/13, •••), and expanded states (ν=5/8, 7/10, 3/8, 3/10, 4/11, 4/13, 5/13, 5/17, 6/17). Therefore, it is meaningful to calculate all binding energies in the three categories using a single method.[7]

## 2. Total Hamiltonian of FQHS

A two-dimensional electron system in a strong magnetic field $B$ has single-electron eigenstates (Landau states) of the Hamiltonian $H_0$, when neglecting the Coulomb interactions among electrons. We use the coordinates $x$, $y$ and $z$, the directions of which are the current direction $x$, Hall voltage direction $y$, and magnetic field direction $z$. When the confinement potentials for the $y$ and $z$ directions are described as $U(y)$ and $W(z)$, the single electron Hamiltonian is

$$H_0 = (\mathbf{p} + e\mathbf{A})^2/(2m) + U(y) + W(z),$$

where the vector potential is $\mathbf{A} = (-yB, 0, 0)$. The eigenfunction is well known to be the Landau state $\psi_k$, as

$$H_0\psi_k = E_0\psi_k, \quad \psi_k \equiv \psi_k(x,y,z) = u e^{ikx} e^{-\alpha(y-c)^2} \phi(z), \quad \left[-\frac{\hbar^2}{2m}\frac{\partial^2}{\partial z^2} + W(z)\right]\phi(z) = \lambda\phi(z) \quad (1)$$

$$\alpha \approx eB/(2\hbar), \quad b \approx U(c), \quad E_0(k) = \lambda + b + \left[\frac{\hbar^2}{2m}\right]2\alpha, \quad c \approx \frac{k\hbar}{eB}, \quad (2)$$

for the ground states in Landau levels. It is noteworthy that the center-position $c$ of electron in the $y$-direction depends upon the wave number $k$. The total Hamiltonian of many electrons is

$$H_T = \sum_{i=1}^{N} H_0(x_i, y_i, z_i) + \sum_{i=1}^{N-1}\sum_{j>i}^{N} \frac{e^2}{4\pi\varepsilon\sqrt{(x_i - x_j)^2 + (y_i - y_j)^2 + (z_i - z_j)^2}}. \quad (3)$$

The first order wave function for many electrons is

$$\Psi(k_1,\cdots,k_N) = \frac{1}{\sqrt{N!}}\begin{vmatrix} \psi_{k_1}(x_1,y_1,z_1) & \cdots & \psi_{k_1}(x_N,y_N,z_N) \\ \vdots & & \vdots \\ \psi_{k_N}(x_1,y_1,z_1) & \cdots & \psi_{k_N}(x_N,y_N,z_N) \end{vmatrix}. \quad (4)$$

In this case, the sum of the single electron energies and classical Coulomb energies is

$$W(k_1,\cdots,k_N) = \sum_{i=1}^{N} E_0(k_i) + C(k_1,\cdots,k_N), \quad (5)$$

where the following is true.

$$C(k_1,\cdots,k_N) = \int\cdots\int \Psi(k_1,\cdots,k_N)^* \sum_{i=1}^{N-1}\sum_{j>i}^{N} \frac{e^2}{4\pi\varepsilon\sqrt{(x_i-x_j)^2+(y_i-y_j)^2+(z_i-z_j)^2}} \times$$

$$\times \Psi(k_1,\cdots,k_N)\,dx_1 dy_1 dz_1\cdots dx_N dy_N dz_N$$

Consequently, we can divide the total Hamiltonian $H_T$ into the diagonal component $H_D$ and the off-diagonal component $H_I$ as the following (see reference 7).

$$H_D = \sum_{k_1,\cdots,k_N} |\Psi(k_1,\cdots,k_N)\rangle W(k_1,\cdots,k_N)\langle\Psi(k_1,\cdots,k_N)| \tag{6}$$

$$H_I = H_T - H_D \tag{7}$$

### 3. Binding Energy of FQHS

There are many configurations of $N$ electrons occupying the Landau orbitals. The configuration shown in Fig. 1 has minimum energy of the classical Coulomb interactions for the filling factor of 5/8. That is to say, $W$ for Fig. 1 is minimal among all electron configurations with $\nu=5/8$.

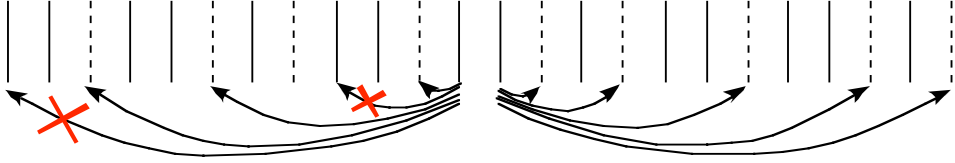

Fig. 1.  Electron configuration for $\nu=5/8$. Dotted lines indicate empty electron orbitals. The arrows show transitions via the interaction Hamiltonian $H_I$.

In this configuration, the number of electron pairs occupying the nearest orbitals is 2 in each unit cell, as is readily apparent in Fig. 1. These electron pairs are more affected by the interaction $H_I$ than the non-nearest pairs. Therefore, we calculate the second-order perturbation energies of $H_I$ for nearest electron pairs. Transitions via $H_I$ should satisfy momentum conservation of the $x$ direction. Therefore, the sum of two wave numbers $k_1$ and $k_2$ before the transition is equal to the sums of $k_1'$ and $k_2'$ after the transition, as $k_1' + k_2' = k_1 + k_2$. The center positions for each electron are $c_1$ and $c_2$ before the transition. They become $c_1'$ and $c_2'$ after the transition (For example: when position $c_1$ is transferred to the sixth orbital to the left, the position $c_2$ is transferred to the sixth orbital to the right through conservation of momentum). We define the integral value $Z_2$ as

$$Z_2 = -\sum_{\Delta k \neq 0, 2\pi/\ell} \frac{\langle k_1, k_2 | H_I | k_1', k_2' \rangle \langle k_1', k_2' | H_I | k_1, k_2 \rangle}{W_G - W_{\text{excite}}(k_1 \to k_1', \, k_2 \to k_2')}, \tag{8}$$

where $W_G$ is the eigenenergy of $H_D$ for the ground state with most uniform electron configuration and $W_{\text{excite}}$ is the excited energy after the transition. Accordingly $W_G < W_{\text{excite}}$. It is noteworthy that the summation in Eq. (8) is carried out for all wave number transfers ($\Delta k = k_1' - k_1$) except 0 and $2\pi/\ell$. The summation may approximate to integration because of the extremely small value of $2\pi/\ell$ for a usual device length. Figure 1 shows that each nearest electron pair can transfer to two orbital pairs, but cannot transfer to one orbital pair within every three vacant orbital pairs. That is to say, two transitions are possible among eight orbitals in each unit cell because all the electron spins have the same direction under a strong magnetic field. Consequently, the binding energy of the nearest electron pair is $(2/8)Z_2$. The two nearest electron pairs exist in a unit cell. The pair energies are shared by five electrons inside a unit cell. Thereby, the binding energy per electron $\omega_2(\nu)$ is $(2/8)Z_2 (2/5)$, namely $\omega_2(5/8) = Z_2/10$. At the filling factor of 3/8, the electron and hole symmetry hold. The electron number is 3 for each unit cell. Therefore, the binding energy per electron $\omega_2(3/8)$ is $(2/8)Z_2 (2/3)$, i.e. $\omega_2(3/8) = Z_2/6$.

We obtain zero for the binding energy per electron with a filling factor of 3/4 because all transitions of nearest electron pair are prohibited as a result of momentum conservation. We show binding energies $\omega_2(\nu)$ in Table I. The binding energy $\omega_2(1/2)$ is zero, because the fractional state with filling factor of 1/2 has no electron pair occupying the nearest neighbor orbitals.

Table I.  Binding energies for fractional filling factors with even denominator

| $\nu$ | $\omega_2(\nu)$ | $\nu$ | $\omega_2(\nu)$ |
|---|---|---|---|
| 1/2 | 0 | | |
| 3/4 | 0 | 1/4 | 0 |
| 5/6 | 0 | 1/6 | 0 |
| 5/8 | $Z_2/10$ | 3/8 | $Z_2/6$ |
| 7/8 | 0 | 1/8 | 0 |
| 7/10 | $2Z_2/35$ | 3/10 | $2Z_2/15$ |
| 9/10 | 0 | 1/10 | 0 |

The calculation value of $Z_2$ was discussed in details in reference 7). The quantum transition number depends upon the forbidden orbital number and allowed orbital number, which are determined by the Pauli exclusion principle. Accordingly the same situation appears in a higher order perturbation calculation of binding energies. We discuss the details in the next section.

4. **Higher Order Calculations**

We can calculate the third order, fourth order, and higher order binding energies through the same method as mentioned in the previous section. We show the third order calculation as an example. A new integral value $Z_3$ is introduced as follows:

$$Z_3 = -\sum_{\Delta k' \neq 0, 2\pi/\ell} \sum_{\Delta k'' \neq 0, 2\pi/\ell} \frac{\langle k_1, k_2 | H_1 | k_1', k_2' \rangle \langle k_1', k_2' | H_1 | k_1'', k_2'' \rangle}{\left(W_G - W_{\text{excite}}(k_1 \to k_1', \ k_2 \to k_2')\right)} \frac{\langle k_1'', k_2'' | H_1 | k_1, k_2 \rangle}{\left(W_G - W_{\text{excite}}(k_1 \to k_1'', \ k_2 \to k_2'')\right)} \quad (9)$$

where $k_1', k_2',$ and $k_1'', k_2''$ express the wave numbers of the intermediate states. They are related to the original wave numbers $k_1, k_2$ as,

$$\begin{aligned} k_1' &= k_1 + \Delta k', \ k_2' = k_2 - \Delta k' \\ k_1'' &= k_1 + \Delta k'', \ k_2'' = k_2 - \Delta k'' \end{aligned} \quad (10)$$

As it is seen in figure 1, each nearest electron pair can transfer to two orbital pairs within every three vacant orbital pairs. Two transitions are possible among eight orbitals in each unit cell. Accordingly, the number of allowed momentum transfers $\Delta k'$ and $\Delta k''$ are 2/8 of all transfers in the summation (9). This ratio $g$ is described as $g = 2/8$ at $v = 5/8$. Therefore, the third order binding energy of the nearest electron pair for $v = 5/8$ is

$$g^2 Z_3 = (2/8)^2 Z_3 = Z_3/16 \quad (11)$$

The two nearest electron pairs exist in a unit cell as in figure 1. Thereby, the third order binding energy per electron $\omega_3(v)$ is

$$\omega_3(v) = (2/5) g^2 Z_3 = Z_3/40 \quad \text{for} \ v = 5/8. \quad (12)$$

We show the third order binding energies for the other fractional filling factors in Table II.

Table II.   Third order binding energies per electron

| ν | $\omega_3$ | ν | $\omega_3$ |
|---|---|---|---|
| 1/2 | 0 | | |
| 2/3 | $Z_3/18$ | 1/3 | $Z_3/9$ |
| 3/4 | 0 | 1/4 | 0 |
| 3/5 | $4 Z_3/75$ | 2/5 | $4 Z_3/50$ |
| 4/5 | $Z_3/100$ | 1/5 | $Z_3/25$ |
| 5/6 | 0 | 1/6 | 0 |
| 4/7 | $9 Z_3/196$ | 3/7 | $9 Z_3/147$ |
| 5/7 | $4 Z_3/245$ | 2/7 | $4 Z_3/98$ |
| 6/7 | $Z_3/294$ | 1/7 | $Z_3/49$ |
| 5/8 | $Z_3/40$ | 3/8 | $Z_3/24$ |
| 7/8 | 0 | 1/8 | 0 |
| 5/9 | $16 Z_3/441$ | 4/9 | $16 Z_3/324$ |
| 7/9 | $4 Z_3/567$ | 2/9 | $4 Z_3/162$ |
| 8/9 | $Z_3/648$ | 1/9 | $Z_3/81$ |
| 7/10 | $2 Z_3/175$ | 3/10 | $2 Z_3/75$ |
| 9/10 | 0 | 1/10 | 0 |

The higher order binding energies can be calculated through a similar method. The higher order contribution becomes quite small for small value of $g$, because the factor $g^n$ is multiplied to $Z_n$ in $n$-th order calculation.

## 5. Conclusion

The calculation of this paper show that the binding energies are non zero values for specific fractional filling numbers with even denominators (for example ν=3/8, 5/8, 3/10, 7/10). Therefore these fractional quantum Hall states with even denominator are stable. As is shown in Table I, the value of the binding energy takes the largest magnitude for ν=3/8, and takes the smallest magnitude for ν=7/10 among ν=5/8, 3/8, 7/10 and 3/10. The experimental data also show that the depth in the dip of $\rho_{xx}$ is deepest at ν=3/8 and most shoal at ν=7/10 among ν=5/8, 3/8, 7/10 and 3/10.[1] Consequently, our theoretical result is in agreement with the experimental data.

The present calculation method of binding energies can be applied to all fractional filling factors with even denominator or odd denominator. (In the traditional theory, the FQH states with even denominator cannot be understood using the standard composite fermion model.) Therefore, the present method is useful to examine all states with fractional filling factors.

**Acknowledgment**

The author thanks Prof. K. Terakura (Regional Editor of Surface Science) for recommendation to submit this manuscript to Journal of the Physical Society of Japan.

1) W. Pan, H. L. Stormer, D. C. Tsui, L. N. Pfeiffer, K. W. Baldwin, and K. W. West, Phys. Rev. Lett. **88**, (2002) 176802.
2) H. L. Stormer, *Nobel lectures, Physics 1996–2000*, (World Scientific), pp. 321.
3) J. K. Jain, Phys. Rev. Lett. 62 (1989) 199; Phys. Rev. B 41 (1990) 7653.
4) A. Wojs and J. J. Quinn, Phys. Rev. B 61, 2846 (2000); A. Wojs, K.-S.Yi, and J. J. Quinn, Phys. Rev. B 69, (2004) 205322.
5) J. H. Smet, Nature (London) 422, (2003) 391; M. R. Peterson and J. K. Jain cond-mat/0309291.
6) E. A. Pashitskii, Low Temp. Phys. 31, (2005) 171.
7) S. Sasaki, Physica B **281** (2000) 838; S. Sasaki, *Proc. 25th Int. Conf. Phys. Semicond.*, (Springer 2001) 925; S. Sasaki, Surface Science **532–535** (2003) 567; S. Sasaki, Surface Science **566–568** (2004) 1040; S. Sasaki, *Surface Science: New Research* (Nova Science Publishers 2006) Chap. 4, p. 103-161.